\documentclass[9pt,twocolumn,twoside]{opticajnl}
\usepackage[utf8]{inputenc}
\usepackage{amsmath} 

\journal{opticajournal} 
\setboolean{shortarticle}{true}

\title{Fiber-based diffractive deep neural network}

\author[1]{Bahadır Utku Kesgin}
\author[2]{Firdevs Yüce}
\author[1,*]{Uğur Teğin}

\affil[1]{Koç University Department of Electrical and Electronics Engineering, İstanbul, Türkiye}
\affil[2]{Koç University Department of Mechanical Engineering, İstanbul, Türkiye}

\affil[*]{utegin@ku.edu.tr}

\begin{abstract}
Optical computing has reemerged as a promising alternative computing paradigm for providing energy-efficient information processing in the age of artificial intelligence. Among various photonic neural network platforms, diffractive optical processing systems in free space proved high-performance computing with high parallelism. Here, we report fiber-based diffractive deep neural networks by optimizing the linear coupling of the waveguide modes. Our approach demonstrated high performance in various machine learning tasks such as biomedical disease, fashion, and geospatial classification with a simple readout layer and all-optically. Operating on linear optics, our architecture performs on par with neural networks even in complex datasets where the data cannot be separated using linear operations. These results will enable efficient and scalable diffractive information processing with waveguides for real-life computing, telecommunications, and imaging applications.
\end{abstract}

\setboolean{displaycopyright}{false} 

\begin{document}

\maketitle
In recent years, architecture-limited conventional computing platforms have begun to struggle with power consumption when computing large data sets for artificial neural networks \cite{strubell_energy_2020}. The emerging need for a new high-speed and energy-efficient computing paradigm with high parallelism has led to the search for new computing schemes in different domains. Optical computing has long been envisioned as a transformative approach to overcoming these fundamental limitations of conventional computing methods\cite{mcaulay1991optical}. Using light for information processing, optical systems inherently offer massive parallelism, high bandwidth, and energy efficiency, advantages that are particularly critical in the era of data-intensive computing \cite{McMahon2023-ak, Wetzstein2020-pd}. Early optical computing efforts focused on performing fundamental analog operations, such as Fourier transforms and convolution, finding applications in signal processing and simple neural network architectures such as Hopfield networks\cite{lugt1974coherent, Farhat1985-zr, goodman2005introduction}. 

Recently, optical computing, particularly photonic neural networks, has regained interest in overcoming the extensive challenges of conventional hardware for machine learning applications. If all-optical computing is not possible or hard to achieve, photonic neural networks use optical information processing via light propagation to extract features from samples of machine learning datasets. Thus, the complex machine learning task becomes solvable with simple methods applied at the readout layer or all-optically. Various promising architectures have been proposed with free-space propagation \cite{pierangeli2019large,pierangeli2021photonic}, complex random media \cite{rafayelyan2020large,wang2024large,Carpinlioglu2024-nc, shen2017deep}, optical nonlinearities \cite{Tegin2021-ao, Muda2024-yh, Oguz2024-pd, Kesgin2025-sx, oguz_training_2025}, and repetition-based nonlinear encoding \cite{Yildirim2024-cv, Xia2024-gf}.  

To enhance reconfigurability in linear optical processors, diffraction is used as an information processor through computationally optimized diffractive plates called Diffractive Deep Neural Networks (D\textsuperscript{2}NNs) \cite{Lin2018-sq}. Such an optical computing architecture provides enormous degrees of freedom to program optical diffraction with numerical optimization and 3D printing. At terahertz wavelength due to the resolution of 3D printing technology and offers a compact free space analog computing platform. D\textsuperscript{2}NNs are used for a variety of applications such as all-optical computing \cite{Lin2018-sq}, sub-wavelength imaging \cite{Hu2024-hs}, phase conjugation \cite{Shen2024-hd}, and secure free-space optical communications \cite{Yang2024-iu}. 

Multimode fiber acts as a relatively long diffusing media when the propagation is linear with low power light. In such a condition, linear coupling between the waveguide modes become dominant effect due to fabrication imperfections and bending of the fiber. Controlling light propagation in multimode fibers via mechanical perturbations is possible and it has been employed for various applications such as nonlinear frequency generation \cite{Qiu2024-hf}, shape single photons \cite{Shekel2023-xs} wavefront shaping using a transmission matrix or black-box optimization \cite{Resisi2020-bx, Marima2025-mm}. On the other hand, fiber-based controllable implementations of linear optical computing offer significant advantages in power consumption and seamless integration with existing optical communication and imaging platforms. Despite these benefits, fiber-based diffractive optical computing remains largely unexplored.

Here, we present a fiber-based diffractive deep neural network (Fiber-D\textsuperscript{2}NN) architecture that leverages controllable linear propagation within a multimode fiber as a computation medium. Our approach optimizes linear coupling of the waveguide modes through mechanical perturbations to achieve high-performance and scalable optical computing. Fiber-D\textsuperscript{2}NN experimentally tested on complex datasets for machine learning tasks and performs on par with contemporary neural networks. We experimentally verify that through optimization of the linear coupling of the modes, Fiber-D\textsuperscript{2}NNs perform high-performance feature extraction, which helps to separate datasets' classes via applying data analysis techniques to the optically processed information. The proposed Fiber-D\textsuperscript{2}NN architecture represents a step toward integrating diffractive optical computing with fiber-optic networks, enabling information processing directly within existing optical interconnects.

\begin{figure}[t]
\centering
\includegraphics[width=\linewidth]{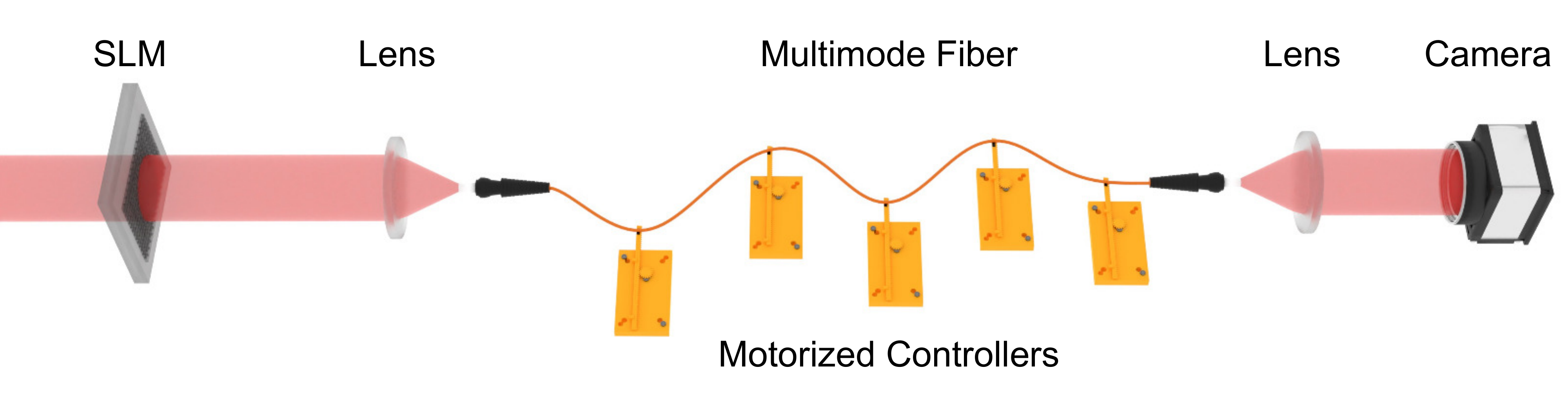}
\caption{Experimental setup. SLM, Spatial light modulator.}
\label{setup}
\end{figure}

Figure \ref{setup} illustrates the experimental setup of Fiber-D\textsuperscript{2}NN. We use a continuous wave laser centered at 1064 nm with a Gaussian beam profile as the light source.
The data to be optically processed is encoded on the Gaussian beam as a phase modulation with a phase-only spatial light modulator (SLM) (Holoeye Pluto-2.1 NIR-145). The spatially modulated beam is focused on the input facet of the graded-index multimode fiber through a lens. A commercially available graded-index multimode fiber with a 1-meter length, 62.5 $\mu$m core radius, and a numerical aperture (NA) of 0.275 is used in our experiments. The fiber supports up to 344 spatial modes at the operation wavelength for a single polarization. At the distal facet of the multimode fiber, the optically processed data is captured using a camera with a pixel pitch of 7.84 $\mu$m as a far field image (see Supplementary Note 1).

To optimize the linear mode coupling of the multimode fiber we place 9 controllers to perturb the fiber orthogonally to the propagation axis. Controllers feature a stepper motor (28BYJ-48) which is connected to a gear that translates radial motion to linear motion which is transferred to movement sticks that move the fiber forward or backward. Changes in the lateral position of the control points create induced mode coupling that changes the overall diffractive optical information processing inside the fiber. The magnitude of the corresponding change is controlled by the rotation step size of the motors. To find the optimal controller steps for each task, we utilize the Trust Region Bayesian Optimization \cite{eriksson_scalable_2019} algorithm, which enables sample-efficient optimization to program our optical computing scheme (see Supplementary Notes 1 and 2).

During the optimization process of Fiber-D\textsuperscript{2}NN, in each iteration, the values given by the Bayesian optimizer are applied to the controllers; thus, the coupling between the fiber modes is changed. Later, the selected datasets are optically processed, and the camera records the resulting light distributions. At the digital readout layer, average pooling and a simple ridge classifier are performed on the recorded optically processed data to calculate the accuracy of the machine-learning task. The achieved performance is yielded back to the optimizer to be increased. The use of ridge classification prevents overfitting and is commonly used in analog computing architectures. In this study, all of the datasets are divided with a ratio of 80\% training set and 20\% test set (see Supplementary Note 2).

\begin{figure}[t]
\centering
\includegraphics[width=0.8\linewidth]{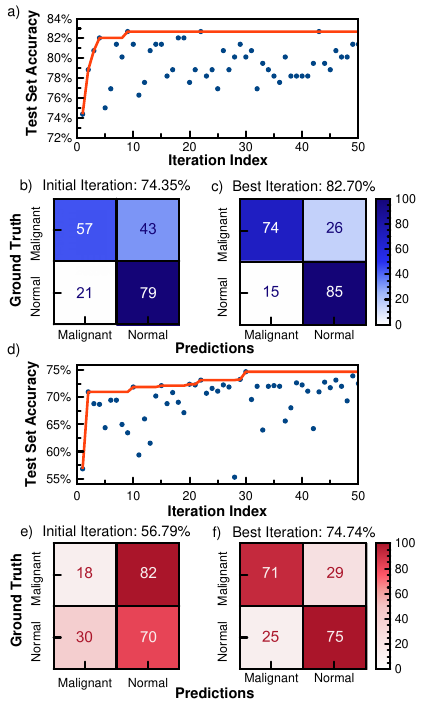}
\caption{Results of all-optical and ridge classification on BreastMNIST dataset. Evolution of classification accuracies for 50 iterations in ridge (a) and all-optical (d) classification. Classification confusion matrix of the initial optimization iteration for ridge (b) and all-optical (e) classification. Classification confusion matrix of the final optimization iteration for ridge (c) and all-optical (f) classification.}
\label{fig2}
\end{figure}
We first test the performance of Fiber-D\textsuperscript{2}NN, with a binary biomedical classification task using Breast MNIST dataset \cite{Yang2023-xf}. Here, Fiber-D\textsuperscript{2}NN decides whether a tumor is malignant or benign for breast ultrasound images. Without any optical processing, our digital readout layer results in 75\% accuracy, and it serves as a baseline for understanding the impact of optical computing architecture. For 50 iterations, the Bayesian optimizer optimizes the linear mode coupling in the multimode fiber by changing the mechanical perturbation conditions to increase the Fiber-D\textsuperscript{2}NN's performance. The optimizer increases the classification accuracy from 74.35\% to 82.70\%, as illustrated in Figure \ref{fig2}(a). As a comparison, a conventional artificial neural network (ResNet-18) with 11 million parameters can only achieve 83.30\% accuracy for this relatively challenging binary classification task.

After the promising results of the Fiber-D\textsuperscript{2}NN with a readout layer, we study all-optical inference with the same dataset to test the performance of our approach. We divide the obtained camera image radially by the number of classes for all-optical classification. We assign each class a section and sum the intensities in each section. The class associated with the section that has the highest intensity is selected as the final decision. As shown in Figure \ref{fig2}(d), the optimizer increases overall accuracy from 56.79\% to 74.74\% after 50 iterations of optimization. Such an increase in accuracy shows the impact of mode coupling optimization on the fiber-based diffractive network in all-optical inference.
\begin{figure}[t]
\centering
\includegraphics[width=0.82\linewidth]{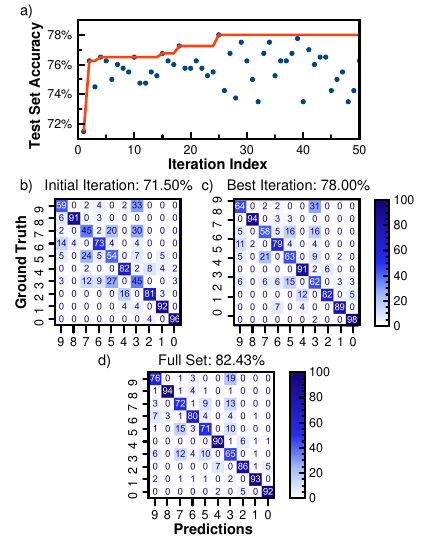}
\caption{Results of ridge classification on FashionMNIST dataset. (a) Evolution of subset classification accuracies for 50 iterations in ridge classification. Subset classification confusion matrix of the initial (b) and final (c) optimization iteration for ridge classification. (d) Full set classification confusion matrix of the final optimization iteration for ridge classification. }
\label{fig3}
\end{figure}

Encouraged by the success of Fiber-D\textsuperscript{2}NN on binary classification task, we test the performance of our optical computing architecture with challenging multi-class datasets. For this goal, we first select Fashion MNIST\cite{Xiao2017-hi}, which is commonly used in the field of optical computing. The dataset is composed of 70,000 samples from 10 classes of fashion items. Due to a high number of samples, to decrease the optimization time in experiments, we subset the dataset by randomly selecting 200 samples from each of the 10 classes and applying the optimization on the subset accuracy. After finding the optimum conditions for the mode coupling to achieve the best test accuracy on the subset, we optically process the entire dataset at this optimal coupling point. Without any optical processing, the digital readout layer results in 64.5\% accuracy for the subset and 75.82\% accuracy for the entire Fashion MNIST dataset. During the optimization process of the Fiber-D\textsuperscript{2}NN test, the accuracy of the subset is increased from 71.50\% to 78\%, as illustrated in Figure \ref{fig3}. When the entire dataset is tested for the optimized Fiber-D\textsuperscript{2}NN, we obtain an accuracy of 82.43\%, demonstrating the effectiveness of the Fiber-D\textsuperscript{2}NN architecture in multi-class datasets.

\begin{figure}[t]
\centering
\includegraphics[width=\linewidth]{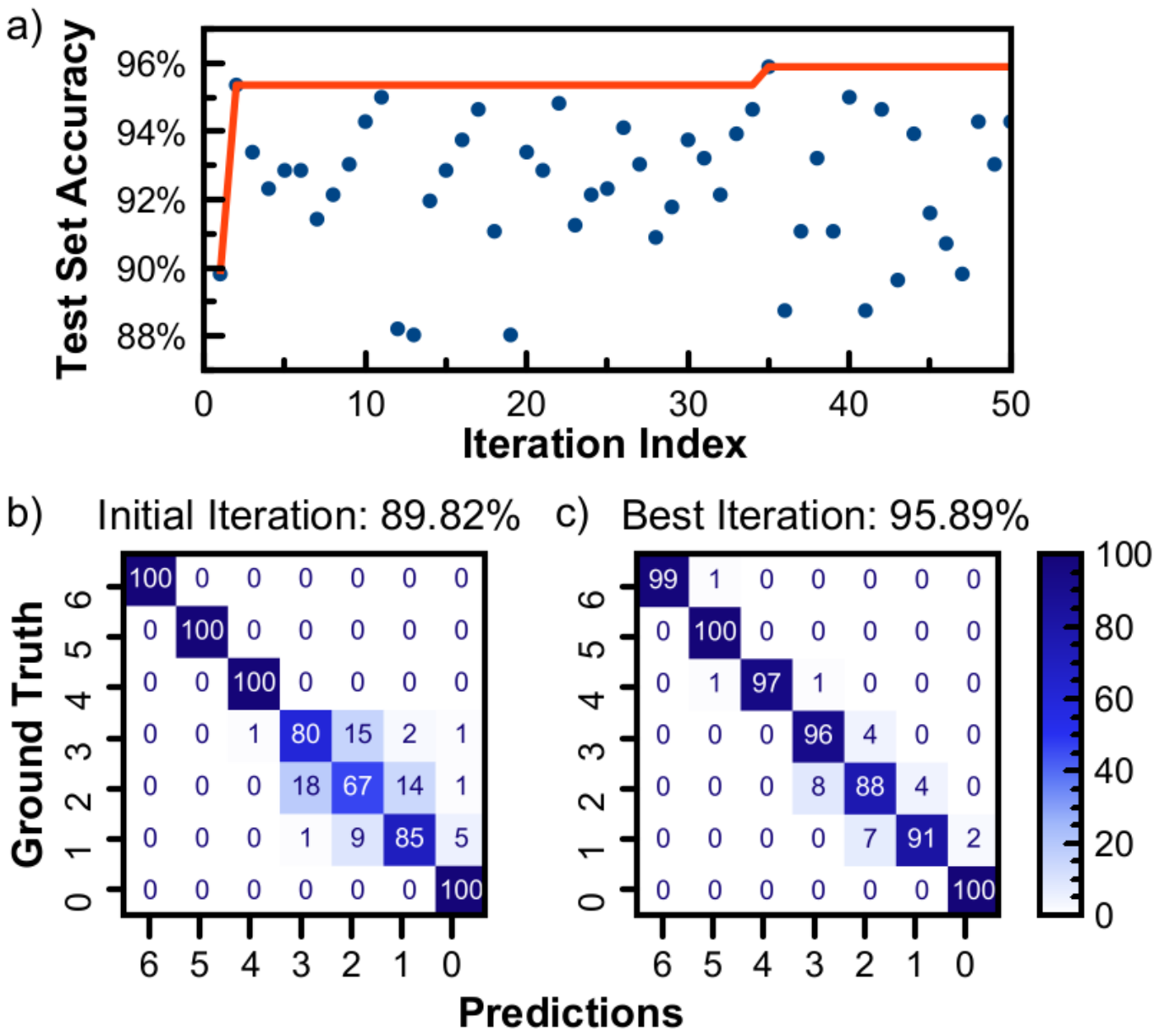}
\caption{Results of ridge classification on RSSCN7 dataset. (a) Evolution of classification accuracies for 50 iterations in ridge classification. Classification confusion matrix of the initial (b) and final (c) optimization iteration for ridge classification.}
\label{fig4}
\end{figure}

While we obtain encouraging results in multi-class classification datasets, the Fashion MNIST dataset is far from having real-world applications. It only serves as a benchmark dataset in the machine learning field. Therefore, we study Fiber-D\textsuperscript{2}NN with another multi-class dataset, satellite-based geospatial classification dataset RSSCN7 \cite{Zou2015-xl}. RSSCN7 features 2800 samples of 7 classes of satellite imagery, which is challenging even for the well-known deep learning models in the literature \cite{Sun2021-ea}. Thus, without any optical processing, the digital readout layer of our architecture (the ridge classifier) results in 18.94\% test accuracy. This shows that classes of this dataset cannot be separated using a single readout layer. Similar to the previous datasets, we employ Bayesian optimization on the Fiber-D\textsuperscript{2}NN, this time for the RSSCN7 with 50 iterations. As shown in Figure \ref{fig4}, the optimizer increases overall accuracy from 89.82\% to 95.90\%. This remarkable test accuracy of Fiber-D\textsuperscript{2}NN for this task is on par with well-known deep neural networks, as demonstrated in Table \ref{table1}. These results demonstrate the effectiveness and generalizability of task-specific trained Fiber-D\textsuperscript{2}NNs for highly complex datasets.
\begin{table}[h]
\centering
\caption{\bf Comparison of different machine learning models in RSSCN7 dataset}
\begin{tabular}{cccc}
\hline Learner & Parameter Count & Test Accuracy$^\textit{a}$\\
\hline
\textit{\textbf{Proposed}} & \textit{\textbf{3609}} & \textit{\textbf{93.57\%}}\\
ResNet-50 & $26,000,000$ & $93.64\%$ \\
VGG16 & $138,000,000$  &$93.57\%$ \\
AlexNet & $60,000,000$  &$91.85\%$ \\
\hline
\end{tabular}
  \label{table1}
  
$^\textit{a}$Training and test sets are split 50\%-50\% for comparison. Benchmark accuracies are obtained from \cite{Sun2021-ea}.
\end{table}

\begin{figure}[t]
\centering
\includegraphics[width=0.8\linewidth]{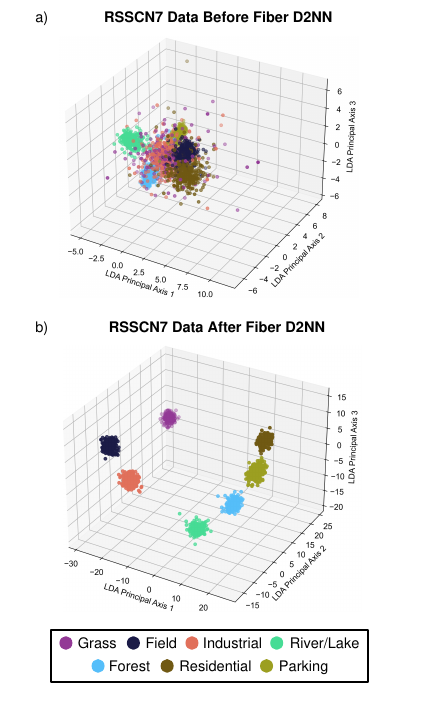}
\caption{Impact of optimized mode coupling and feature extraction with Fiber-D\textsuperscript{2}NN on RSSCN7 data points. (a) Feature space before chaotic transformation. (b) Feature space after chaotic transformation.}
\label{fig5}
\end{figure}

Further investigation is performed to understand the overall impact of optical computing with Fiber-D\textsuperscript{2}NN on the machine learning datasets. We perform dimensionality reductions using Linear Discriminant Analysis (LDA) to visualize the feature space of the dataset with and without optical computing and see the benefit of Fiber-D\textsuperscript{2}NN on input data. The feature space calculated via LDA represents the perspective of the readout layer, which draws the decision lines and thus performs the desired classification tasks. The structure of the raw dataset before the optical computing can be seen from the initial distribution shown in Figure \ref{fig5} (a). It is obvious for this initial distribution that a simple linear readout function cannot successfully separate the classes of this dataset. Such a complex distribution requires nonlinear functions or computing to be separated. In the absence of strong nonlinear functions like Kerr nonlinearity of the fibers \cite{Tegin2021-ao, Muda2024-yh, Oguz2024-pd, Kesgin2025-sx, oguz_training_2025}, a tunable linear optical computing platform such as our Fiber-D\textsuperscript{2}NN architecture can be useful. Figure \ref{fig5} (b) demonstrates the feature space for the same dataset after being processed by the optimized Fiber-D\textsuperscript{2}NN. It shows that Fiber-D\textsuperscript{2}NN can properly map input data to a space where separating them with the simple readout layer is feasible. After Fiber-D\textsuperscript{2}NN, data points from the same classes group together, making the data easily separable and achieving significantly higher accuracy. This results indicates that Fiber-D\textsuperscript{2}NN extracts features and maps data as effectively as nonlinear functions while relaying on linear configurable optics.

Fiber-D\textsuperscript{2}NN yields machine learning performance on par with modern neural networks while consuming 50W on training and 30 W on the testing process. Furthermore, Fiber-D\textsuperscript{2}NN is scalable with fiber modes, which can be increased by decreasing the laser wavelength or increasing the fiber core radius. With commercially available SLMs and optics, Fiber-D\textsuperscript{2}NN can be scaled to compute 4K (4160 x 2464 pixel) data without needing more power. High-resolution optical computing with low power consumption provides a promising edge computing platform to be deployed in real-world applications.

In conclusion, we demonstrate a fiber-based diffractive deep neural network by mechanically optimizing the linear mode coupling within a graded-index multimode fiber. Our results show that fiber-based diffractive information processing can accurately infer from linearly inseparable data using linear optical relations. Our optical computing architecture paves the way for creating scalable and environmentally friendly computer technology. We believe the reported results are of great interest to optical communication, optical imaging and analog computing.

\begin{backmatter}
\bmsection{Funding} This work is supported by the Scientific and Technological Research Council of Turkey (TÜBİTAK) under grant number 122C150.

\bmsection{Disclosures} The authors declare no conflicts of interest.

\bmsection{Data availability} Data underlying the results presented in this paper are not publicly available at this time but may be obtained from the authors upon reasonable request.

\bmsection{Supplemental document}
See Supplement 1 for supporting content.
\end{backmatter}
\bibliography{references}

\end{document}